\documentclass[a4paper,12pt]{article}
\usepackage{amsmath,amsfonts,amssymb,amscd,enumerate}

\def\A{\mathcal{A}}

\def\p{\partial}

\def\U{\mathcal{U}}
\def\P{\mathcal{P}}
\def\D{\mathcal{D}}
\def\x{\hat x}

\begin{document}
\begin{center}
{\Large \bf Kappa Snyder deformations of Minkowski spacetime,
  realizations and Hopf algebra}
\end{center}
\bigskip

\begin{center}

S. Meljanac {\footnote{e-mail: meljanac@irb.hr}},
  D. Meljanac \footnote{e-mail: dmeljan@irb.hr},
 A. Samsarov {\footnote{e-mail: asamsarov@irb.hr}}
and M. Stoji\'c {\footnote{e-mail: marko.stojic@zg.htnet.hr}} \\
 Rudjer Bo\v{s}kovi\'c Institute, Bijeni\v cka  c.54, HR-10002 Zagreb,
Croatia \\[3mm]

\end{center}
\setcounter{page}{1}
\bigskip

\begin{abstract}
       We present Lie-algebraic deformations of Minkowski space with
       undeformed Poincar\'{e} algebra. These deformations interpolate
       between Snyder and $\kappa$-Minkowski space. We find
       realizations of noncommutative coordinates in terms of
       commutative coordinates and derivatives. By introducing
       modules, it is shown that although deformed and undeformed
       structures are not isomorphic at the level of vector spaces,
       they are however isomorphic at the level of Hopf algebraic
       action on corresponding modules.
      Invariants and tensors
       with respect to Lorentz algebra are discussed. A general
       mapping from $\kappa$-deformed Snyder to Snyder space is
       constructed. Deformed Leibniz rule, the Hopf structure and star
       product are found. Special cases, particularly Snyder and
       $\kappa$-Minkowski in Maggiore-type realizations are discussed.
       The same generalized Hopf
       algebraic structures are as well considered in the case of an arbitrary
       allowable kind of realisation and results are given
       perturbatively up to second order in deformation parameters.
\end{abstract}

\section{Introduction}
    Current progress in high energy physics in considerable part very
    much relies on concepts and ideas which come from the scope of
    noncommutative (NC) physics. These concepts embody a modification
    in a description of spacetime as understood from the point of view
    of standard commutative field theories, where it is considered as
    a nondiscrete continuum, i.e. continuous, nondiscretized geometric structure.
  A signal for the possible modification of spacetime structure
    emerges through the appearance of a new fundamental length
    scale in physics. This new fundamental length scale, known as Planck
    length \cite{Doplicher:1994zv},\cite{Doplicher:1994tu}, appears
    within two different and highly significant physical contexts.
    The first one comes from loop quantum gravity in
    which the Planck length plays a fundamental role. There,
    a process of quantization leads to the area and volume operators
    having discrete spectra,
    with minimal values of the corresponding eigenvalues being proportional to the square
    and cube of Planck length, respectively \cite{Rovelli:2004tv},\cite{Thiemann:2002nj}.
 The second important context where the signal for the existence of a
    new fundamental length scale can be found is related to 
    certain observations of ultra-high energy cosmic rays
    which seem to contradict the usual understanding of some
    astrophysical processes like, for example, electron-positron
    production in collisions of high energy photons. It turns out that
    deviations observed in these processes can be explained by
    modifying dispersion relation in such a way as to incorporate the
    fundamental length scale \cite{AmelinoCamelia:1997gz},\cite{AmelinoCamelia:2000zs},\cite{Gambini:1998it},\cite{Alfaro:1999wd}.
   The appearance of this new fundamental length scale has a far
    reaching consequence for the spacetime structure because at this
    scale the spacetime structure becomes
    discretized and fuzzy and thus most conveniently described in
    terms of noncommutative geometry.
   NC spacetime has also been revived in
    the paper by Seiberg and Witten \cite{Seiberg:1999vs} where NC manifold emerged in a
    certain low energy limit of open strings moving in the background
    of a two form gauge field. Recently, a $\kappa$-Minkowski NC space
  in bicrossproduct basis was shown to emerge from
  consideration of a NC differential structure on the
  (pseudo)-Riemannian manifold \cite{Majid},\cite{Majid:2010mj}.

  As a result of different approaches to quantum gravity, various
  phenomenological models emerged, whose aim is to predict a
  phenomenology that tests and searches for specific properties the
  fundamental theory of quantum gravity might have. Among them the
  most interesting are Lorentz invariance violation (LIV) models \cite{Jacobson:2005bg},\cite{AmelinoCamelia:2008qg} and
  the models of the so called doubly special relativity theory (DSR)
 \cite{AmelinoCamelia:2000mn},\cite{AmelinoCamelia:2000ge},\cite{Magueijo:2001cr}. 
  In LIV models the observer independence is explicitly violated and a
  preffered frame is singled out. Preffered frame in this case is
  typically thought to be the rest frame of the cosmic microwave
  background radiation. On the other side, in DSR models postulates of
  relativity may be reformulated so as to avoid a necessity for
  singling out a preffered frame. It is this set of models that
  provide a kinematical theory within which a Planck length is
  incorporated as a new fundamental invariant, besides that of the
  speed of light $c$.
 Currently there is no unique view about the source where doubly
  special relativity could possibly originate from.
While the motivation for DSR comes originally from loop quantum
  gravity, there are some arguments suggesting that it may as well emerge in a
  form of a theory described in terms of what is known as $\kappa$-Poincar\'{e}
  algebra, after taking an appropriate low energy limit of quantum gravity.
Most of considerations on DSR has been made within the framework of
$\kappa$-Poincar\'{e} algebra, a type of quantum Hopf algebra where
  the algebraic symmetry properties are considered
 as necessarily emerging from a deformation of Poincar\'{e}
 and even Lorentz symmetry 
 \cite{Lukierski:1991pn},\cite{Lukierski:1993wx},\cite{Majid:1994cy},\cite{zakrzewski},\cite{KowalskiGlikman:2002we},\cite{KowalskiGlikman:2002jr}.
In some recent considerations it is indicated that a doubly special
  relativity framework does not necessarily require a deformation of
  the Lorentz group, neither its action.
  As example, the minimal canonical covariantisation of the usual
  $\kappa$-Minkowski model was shown in \cite{Dabrowski:2009mw} to
  give rise to a covariant 
 $\kappa$-Minkowski spacetime, preserving Lorentz symmetry.
  Another analysis which goes along the similar track 
 was put forward
  in \cite{Meljanac:2010ps} by requiring the parameter of deformation to
  transform as a
  vector under Lorentz generators.
Certain aspects of DSR models are however currently under debate
  \cite{Hossenfelder:2009fc},\cite{Hossenfelder:2010tm},\cite{Jacob:2010vr}. 
   
In this paper, we shall be interested in still wider class of deformations
 of $\kappa$-Minkowski space that can be described within a generalization of
 $\kappa$-Poincar\'{e} algebra. $\kappa$-Poincar\'{e} algebra alone is an algebra that describes in a
 direct way only the energy-momentun sector of the physical
 theory. This means that this algebra is specified by commutation relations
 between linear and angular momenta only.
 Although this sector alone is insufficient to set up physical
 theory, the Hopf algebra structure makes it possible to extend the
 energy-momentum algebra to the algebra of spacetime. 

 It is known \cite{KowalskiGlikman:2002jr} that there exists a
 transformation which maps $\kappa$-Minkowski spacetime into
 spacetime with noncommutative structure described by the algebra
 first introduced by Snyder \cite{snyder}. Although this map is not an
 isomorphism,
Snyder algebra itself is particularly interesting since it is
 compatible with Lorentz
 symmetry  and  it also provides
 \cite{KowalskiGlikman:2002jr} configuration space
 consistent with DSR and thus can be used
 to construct the second order  particle
 Lagrangian. The last observation makes possible to define physical
 four-momenta determined by the particle dynamics. This would be
 significant step forward because the theoretical development in this
 area has been mainly kinematical so far.
 One such dynamical picture
 has been given recently in \cite{Ghosh:2006cb} where it was shown that
 reparametrisation symmetry of the proposed Lagrangian, together with
the appropriate change of variables and conveniently
 chosen gauge fixing conditions, leads to an algebra which is a
 combination of $\kappa$-Minkowski and Snyder algebra. This
 generalized type of algebra describing noncommutative structure of
 Minkowski spacetime is shown to be consistent with the
 Magueijo-Smolin dispersion relation \cite{Magueijo:2001cr}. This type of NC space is also
 considered in \cite{Chatterjee:2008bp}. It has to be stressed that NC spaces
 in neither of these papers are of Lie-algebra type.

   In order to fill this gap, in the present paper we unify
  $\kappa$-Minkowski and Snyder space in a more general NC space
  which is of a Lie-algebra type and, in addition, is
    characterized by the
   undeformed Poincar\'{e} algebra and deformed coalgebra.
   In other words, we shall consider a type of NC space which
  interpolates between
 $\kappa$-Minkowski space and Snyder space in a Lie-algebraic way and has
    all deformations contained within the coalgebraic sector.
  First such example of NC space with undeformed Poincar\'{e} algebra, but with deformed coalgebra
    is given by Snyder \cite{snyder}. Some other types of NC spaces are
   also considered within the approach in which the Poincar\'{e}
   algebra is undeformed and coalgebra deformed, in particular the
   type of NC space with $\kappa$-deformation \cite{KowalskiGlikman:2002we},\cite{KowalskiGlikman:2002jr},\cite{Meljanac:2006ui},\cite{KresicJuric:2007nh},\cite{Meljanac:2007xb},\cite{Borowiec:2009vb},\cite{Borowiec:2009ty}.
  Here we present a broad class of Lie-algebra type deformations with the same
   property of having deformed coalgebra, but undeformed algebra.
    The investigations carried out in this paper  are along the track
  of developing general techiques of calculations,
              applicable for a widest possible class of NC spaces and
 as such  are a continuation of the work done in a series of previous
   papers
 \cite{Meljanac:2006ui},\cite{KresicJuric:2007nh},\cite{Meljanac:2007xb},\cite{Jonke:2002kb},\cite{Durov:2006iv},\cite{Meljanac:2008ud},\cite{Meljanac:2008pn}.
  They are in particular a continuation of Snyder space analysis
  undertaken in \cite{Battisti:2008xy},\cite{Battisti:2010sr}.
 The methods used in these investigations were taken over from the
  Fock space analysis carried out in \cite{Doresic:1994zz},\cite{bardek}.

   The plan of paper is as follows. In section 2 we introduce a type
   of deformations of Minkowski space that have a structure of a Lie
   algebra and which interpolate between $\kappa$-type of deformations and
   deformations of the Snyder type.
      In section 3 we  analyze realizations of NC space in terms
   of operators belonging to the undeformed Heisenberg-Weyl algebra.
   The introduction of modules in section 4 makes it possible to
   establish an isomorphism between deformed and undeformed
   structures, which otherwise does not exist. This is due to the fact
   that in the case of kappa-Snyder type of deformation, deformed algebra of noncommutative functions is much larger
   than the corresponding undeformed algebra of commutative functions,
   making any isomorphism between these two impossible.
   Described situation can however be overcome if one introduces
   module for the envelopes of the deformed and undeformed Heisenberg
   algebras, respectively, and looks at their action on
   the unit element in the module.
  Then corresponding projections appear to be isomorphic to each other.
 In section 5 we tackle the
   issue of the way in which general invariants and tensors can be
   constructed out of NC coordinates.
 Section 6 is devoted to an analysis of the effects which these
   deformations have on the Hopf structure of the symmetry algebra and after that, in
   section 7 we
   specialize the general results obtained to some interesting special
   cases,  such as $\kappa$-Minkowski space and Snyder space.
 In addition, the most general case of realization, consistent with
   all Jacobi identities and given algebra, is analysed
    perturbatively up to second order in deformation parameters
    and resulting coalgebraic structures, such as coproducts and antipodes, are also given up to the same order.
  We end the paper with some discussion.  

\section{Noncommutative coordinates  and Poincar\'{e} algebra}

We are considering a Lie algebra type of noncommutative (NC) space generated by
the coordinates $\x_0, \x_1,\ldots ,\x_{n-1},$ satisfying the commutation
relations
\begin{equation} \label{2.1}
[\x_{\mu},\x_{\nu}]=i(a_{\mu}\x_{\nu}-a_{\nu}\x_{\mu})+s M_{\mu
\nu},
\end{equation}
where indices $\mu,\nu=0,1\dots,n-1$ and $a_0,a_1,\dots ,a_{n-1}$ are componenets
of a $n$-vector $a$ in Minkowski space whose metric signature is
$~ \eta_{\mu\nu} = diag(-1,1,\cdot \cdot \cdot, 1).$ The quantities
$a_{\mu}$ and $s$ are deformation parameters which measure a degree of
deviation from standard commutativity. They are related to length
scale characteristic for distances where it is supposed that noncommutative character of
space-time becomes important. When parameter $s$ is set to zero,
noncommutativity (\ref{2.1}) reduces to covariant version of
$\kappa$-deformation, while in the case that all components of a
$n$-vector $a$ are set to $0,$
 we get the type of NC space considered
for the first time by Snyder \cite{snyder}. The NC space of this type
has been annalyzed in the literature from different points of view \cite{Battisti:2008xy},\cite{Battisti:2010sr},\cite{Guo:2008qp},\cite{Romero:2004er},\cite{Banerjee:2006wf},\cite{Glinka:2008tr},\cite{Yang:2009pm},\cite{Girelli:2009ii},\cite{Girelli:2010wi}.

 The symmetry of
the deformed space (\ref{2.1}) is assumed to be described by an undeformed Poincar\'{e}
algebra, which is generated by
 generators $M_{\mu\nu}$ of the  Lorentz algebra and generators
 $D_{\mu}$ of translations. This means that generators $M_{\mu\nu},~ M_{\mu \nu} = -M_{\nu \mu}, $
 satisfy the standard, undeformed commutation relations,
\begin{equation} \label{2.2a}
[M_{\mu\nu},M_{\lambda\rho}] =
\eta_{\nu\lambda}M_{\mu\rho}-\eta_{\mu\lambda}M_{\nu\rho}
-\eta_{\nu\rho}M_{\mu\lambda}+\eta_{\mu\rho}M_{\nu\lambda},
\end{equation}
with the identical statement as well being true for the generators $D_{\mu},$
\begin{align} \label{2.5}
[D_{\mu},D_{\nu}]&=0,  \\
[M_{\mu\nu},D_{\lambda}]&= \eta_{\nu\lambda}\,
D_{\mu}-\eta_{\mu\lambda}\, D_{\nu}. \label{2.5a}
\end{align}
The undeformed Poincar\'{e} algebra, Eqs.(\ref{2.2a}),(\ref{2.5}) and (\ref{2.5a})
define the energy-momentum sector of the theory considered. However, full
description requires space-time sector as well. Thus, it is of
interest to extend the algebra (\ref{2.2a}),(\ref{2.5}) and (\ref{2.5a})
so as to include NC coordinates $\x_0, \x_1,\ldots ,\x_{n-1},$
and to consider the action of Poincar\'{e} generators on NC
space (\ref{2.1}),
\begin{equation} \label{2.3}
[M_{\mu\nu},\x_{\lambda}]=\x_{\mu}\, \eta_{\nu\lambda}-\x_{\nu}\,
\eta_{\mu\lambda}-i\left(a_{\mu}\, M_{\nu\lambda}-a_{\nu}\,
M_{\mu\lambda}\right).
\end{equation}
The main point is that commutation relations (\ref{2.1}),(\ref{2.2a})
 and (\ref{2.3}) define a Lie algebra generated by Lorentz generators
$M_{\mu\nu}$ and $\x_{\lambda}.$
 The necessary and sufficient condition for consistency of an
extended algebra, which includes generators $M_{\mu\nu}, ~D_{\mu}$ and $\x_{\lambda},$
 is that Jacobi identity holds for all
combinations of the generators $M_{\mu\nu},$ $D_{\mu}$ and $\x_{\lambda}.$
Particularly, the algebra generated by $D_{\mu}$ and $\x_{\nu}$ is a deformed
Heisenberg-Weyl algebra and we require that this algebra has to be of the form,
\begin{equation} \label{2.6}
[D_{\mu},\x_{\nu}] = \Phi_{\mu\nu}(D),
\end{equation}
where $ \Phi_{\mu\nu}(D)$ are functions of generators $D_{\mu},$
which are required to satisfy the boundary condition
$ \Phi_{\mu\nu}(0)=\eta_{\mu\nu}$ and still be consistent with
Eq.(\ref{2.1}) and relevant Jacobi identities, as explained below.
 This condition means that deformed NC
 space, together with the corresponding coordinates, reduces to
 ordinary commutative space in the limiting case of vanishing
 deformation parameters, $a_{\mu}, s \rightarrow 0.$

One certain type of noncommutativity, which interpolates between
Snyder space and $\kappa$-Minkowski space, has already been investigated
 in  \cite{Ghosh:2006cb},\cite{Chatterjee:2008bp} in the context of
 Lagrangian particle dynamics. However, in these papers algebra
 generated by NC coordinates and Lorentz generators is not
 linear and is not closed in the generators of the algebra. Thus, it is not of
Lie-algebra type. In contrast to this, here we consider an algebra
 (\ref{2.1}),(\ref{2.2a}),(\ref{2.3}), which is of
  Lie-algebra type, that is, an algebra
 which is linear in all generators and
 Jacobi identities are satisfied for all combinations of generators of
 the algebra. Besides that, it is important to note that, once having
 relations (\ref{2.1}) and (\ref{2.2a}), there exists only one possible
 choice for the commutation relation between $M_{\mu\nu}$ and $\x_{\lambda},$
which is consistent with Jacobi identities and makes Lie algebra to close,
 and this unique choice is given by the commutation relation (\ref{2.3}).

In subsequent considerations we shall be interested in
possible realizations of the space-time algebra (\ref{2.1}) in terms
of canonical commutative spacetime coordinates $X_{\mu},$
\begin{equation} \label{2.9}
[X_{\mu},X_{\nu}]=0,
\end{equation}
which, in addition, with derivatives
 $D_{\mu} \equiv \frac{\partial}{\partial X^{\mu}}$ close the undeformed Heisenberg algebra,
\begin{equation} \label{2.10}
[D_{\mu},X_{\nu}] = \eta_{\mu\nu}.
\end{equation}
Thus, our aim is  to find a class of covariant $\Phi_{\alpha\mu}(D)$
realizations,
\begin{equation} \label{2.7}
\x_{\mu} =X^\alpha \Phi_{\alpha\mu}(D),
\end{equation}
satisfying the boundary conditions
$~ \Phi_{\alpha\mu}(0)=\eta_{\alpha\mu},~$ and
 commutation relations (\ref{2.1}) and (\ref{2.3}).
In order to complete this task, we introduce the standard coordinate representation
of the Lorentz generators $M_{\mu\nu},$
\begin{equation} \label{2.8}
M_{\mu\nu} = X_{\mu}D_{\nu}- X_{\nu}D_{\mu}.
\end{equation}
All other commutation relations, defining the extended algebra, are then automatically satisfied, as
well as all Jacobi identities among $\x_\mu$, $M_{\mu\nu},$ and
$D_{\mu}.$ This is assured by the construction (\ref{2.7}) and (\ref{2.8}).

   As a final remark in this section, it is interesting to observe that the trilinear
   commutation relation among NC coordinates has a particularly simple form,
\begin{equation} \label{trilinear} 
[[\x_{\mu},\x_{\nu}],\x_{\lambda}]=a_{\lambda}(a_{\mu}\x_{\nu}-a_{\nu}\x_{\mu})+
 s (\x_{\mu}\, \eta_{\nu\lambda}-\x_{\nu}\,
\eta_{\mu\lambda}),
\end{equation}
which shows that on the right hand side Lorentz generators completely
drop out.  Described property of trilinear relations is relevant and
significant only when $s \neq 0$, since otherwise
it makes no difference nor gives any new information. Hence, when $s \neq 0,$ it has important
consequences for the relationship between algebras of commutative and
noncommutative functions, particularly for the properties of the map
that acts between the two. The problem of existing isomorphism is of special
interest in this regard. While, for example, in the case of
$\kappa$-deformation the relationship between deformed and undeformed
algebras of functions is rather clear, the situation when $s \neq 0$
is to a certain extent relatively dim. What lies behind this statement
is that in the case of $\kappa$-deformation, it is easy to establish
the isomorphism between deformed and undeformed algebras, while for
the more general case of $\kappa$-Snyder deformation, when $s \neq 0,$
the situation gets more complicated. It does not mean that in the case
when $s \neq 0$ isomorphism cannot be established. It can,
  but the things are not so easy and straightforward. We shall continue
  to elaborate on these issues in section 4, where we shall show how
  this isomorphism can be constructed.

 To better understand the nature of these mutual relations, it is desirable to
 introduce necessary notions and ingredients.
Thus,
let us define  an enveloping algebra $\hat{\A}_{\x}$ as free algebra generated by
$\x_\mu$ and  divided by the ideal generated with trilinear relations, Eq.(\ref{trilinear}). The algebra
$\hat{\A}_{\x}$ contains unit element $1$.
 We note that in the case when $s \neq 0,$ two monomials $\x_\mu \x_\nu$ and  $\x_\nu \x_\mu,$
 with $\mu\neq \nu,$ are algebraically independent in the algebra
 $\hat{\A}_{\x}$.
Furthermore, if $\mu\neq\nu\neq\lambda\neq\mu $ (mutually
different) among 6 monomials: $\x_\mu\x_\nu\x_\lambda, ~$ 
$ \x_\mu\x_\lambda\x_\nu$,$ ~ \x_\nu\x_\mu\x_\lambda$,$ ~ \x_\nu\x_\lambda\x_\mu$,$~\x_\lambda\x_\mu\x_\nu$ and $\x_\lambda\x_\nu\x_\mu$, there are
two relations:
\begin{equation}
[[\x_{\mu},\x_{\nu}],\x_{\lambda}]=a_{\lambda}(a_{\mu}\x_{\nu}-a_{\nu}\x_{\mu})
+ s (\x_{\mu}\, \eta_{\nu\lambda}-\x_{\nu}\,
\eta_{\mu\lambda}),
\end{equation}
\begin{equation}
[[\x_{\mu},\x_{\lambda}],\x_{\nu}]=a_{\nu}(a_{\mu}\x_{\lambda}-a_{\lambda}\x_{\mu})
 + s (\x_{\mu}\, \eta_{\lambda \nu}-\x_{\lambda}\,
\eta_{\mu\nu}).
\end{equation}
The third relation $[[\x_{\nu},\x_{\lambda}],\x_{\mu}]$ follows
from above two relations by Jacobi identity.
Thus, among these 6 monomials there are only four of them that are
linearly independent.
In special case $\mu=\nu\neq\lambda$ there is only one relation
\begin{equation}
[[\x_{\mu},\x_{\lambda}],\x_{\mu}]=a_{\mu}(a_{\mu}\x_{\lambda}-a_{\lambda}\x_{\mu})-s\x_{\lambda}\eta_{\mu\mu}
\end{equation}
and there are only 2 linearly independent monomials.
We point out that in the case when $s \neq 0$ the algebra
$\hat{\A}_{\x}$ is larger than the algebra
$\A_ X$ generated by commutative generators $X_\mu$ and thus there is
no isomorphism between $\hat{\A}_{\x}$ and $\A_ X$ at the level of vector spaces.

These things will be further elaborated in Section 4, where their
proper understanding will lead to realization of an isomorphism
between deformed and undeformed algebraic structures.
 In the next section we turn to problem of finding an explicit
$\Phi_{\mu\nu}(D)$ realizations (\ref{2.7}).

\section{Realizations of NC coordinates}
Let us define general covariant realizations:
\begin{equation} \label{3.1}
\x_{\mu} = X_{\mu}\varphi + i(aX)\left(D_{\mu}\,
\beta_1+ia_{\mu}D^2\, \beta_2\right)+i(XD)
\left(a_{\mu}\gamma_1+i(a^2 - s) D_{\mu}\, \gamma_2\right),
\end{equation}
where $\varphi$, $\beta_i$ and $\gamma_i$ are functions of
$A=ia_{\alpha}D^{\alpha}$ and $B=(a^2-s)D_{\alpha}D^{\alpha}$. We further
impose the boundary condition that $\varphi(0,0)=1$ as the parameters of
 deformation $a_{\mu} \rightarrow 0$ and
 $s \rightarrow 0.$ In this way we assure that $\x_{\mu}$
reduce to ordinary commutative coordinates in the limit of vanishing deformation.

It can be shown that Eq.(\ref{2.3}) requires the following set of
equations to be satisfied,
$$
\frac{\p\varphi}{\p A}=-1,\qquad \frac{\p\gamma_2}{\p A}=0, \qquad
\beta_1=1, \qquad \beta_2=0, \qquad \gamma_1=0. $$
Besides that, the commutation relation (\ref{2.1}) leads to the additional two equations,
\begin{equation}
\varphi(\frac{\p\varphi}{\p A}+1)=0,
\end{equation}
\begin{equation}
(a^2-s)[2(\varphi+A)\frac{\p\varphi}{\p
B}-\gamma_2(A\frac{\p\varphi}{\p A}+2B\frac{\p\varphi}{\p
B})+\gamma_2 \varphi]-a^2\frac{\p\varphi}{\p A}-s=0.
\end{equation}
The important result of this paper is that
all above required conditions are solved by a general form of
realization which in a compact form can be written as
\begin{equation} \label{3.2}
\x_{\mu}=X_{\mu}(-A+f(B))+i(aX)D_{\mu}-(a^2-s)(XD)D_\mu\gamma_2,
\end{equation}
where $\gamma_{2}$ is necessarily  restricted to be
\begin{equation} \label{3.3}
\gamma_2=-\frac{1+2f(B)\frac{\p f(B)}{\p B}}{f(B)-2B\frac{\p
f(B)}{\p B}}.
\end{equation}
From the above relation we see that $\gamma_{2}$
 is not an independent function, but instead is related
to generally an arbitrary function $ f(B)$, which has to satisfy the boundary condition $f(0)=1$.
Also, it is readily seen from the realization (\ref{3.2}) that
 $~\varphi ~$ in (\ref{3.1}) is given by $~\varphi = -A + f(B).$
Various choices of the function $f(B)$ lead to different realizations
of NC spacetime algebra (\ref{2.1}).
Two particularly interesting cases are
for $f(B)=1$  and  $ f(B)=\sqrt{1-B}$. The later one, when $
 f(B)=\sqrt{1-B},$ will be given special attention since it allows for an
 exact and complete analysis resulting in 
  coproducts and antipodes for the generators of Poincar\'{e} algebra, which are contained in and
 define a Hopf algebraic structure
 in that particular case. Although the exact treatment is reserved for
 the case $ f(B)=\sqrt{1-B}$ only, we shall be interested in other realizations as well
 and, in particular, in the analysis of the most general case, which will be tackled perturbatively.

It is now straightforward to show that the deformed Heisenberg-Weyl
algebra (\ref{2.6}) takes the form
\begin{align} \label{3.4}
[D_{\mu},\x_{\nu}]=\eta_{\mu\nu}(-A+f(B)) +i a_\mu
D_\nu -(a^2-s)D_\mu D_\nu \gamma_2
\end{align}
and that the Lorentz generators $M_{\mu\nu}$ can be expressed in terms of
NC coordinates as
\begin{equation} \label{2.4}
M_{\mu\nu}=(\hat{x}_{\mu}D_{\nu}-\hat{x}_{\nu}D_{\mu})\frac{1}{\varphi}.
\end{equation}

We also point out that in the special case when realization of NC
space (\ref{2.1}) is characterized by the function $
f(B)=\sqrt{1-B}$, there exists a unique  element Z satisfying:
\begin{equation} \label{2.13}
[Z^{-1},\x_{\mu}] = -ia_{\mu} Z^{-1}+sD_{\mu}, \qquad [Z,\x_{\mu}] = ia_{\mu} Z-sD_{\mu}Z^2.
\end{equation}
and also
\begin{equation} \label{2.15}
\x_{\mu}Z\x_{\nu} =\x_{\nu}Z\x_{\mu}, \qquad [Z,D_{\mu}] =0.
\end{equation}
The element Z is  a generalized shift operator \cite{KresicJuric:2007nh} and its expression
in terms of $D_{\mu}$ can be shown to have the form
\begin{equation} \label{16}
  Z^{-1} = -A + \sqrt{1 - B}.
\end{equation}
As a consequence, the Lorentz generators can be expressed in terms of
$Z$ as
\begin{equation} \label{2.17}
M_{\mu\nu}=(\x_{\mu}D_{\nu}-\x_{\nu}D_{\mu})Z,
\end{equation}
and one can also show that the relation
\begin{equation} \label{2.18}
[Z^{-1},M_{\mu\nu}] = -i(a_{\mu}D_{\nu}-a_{\nu}D_{\mu})
\end{equation}
holds.
In what follows we shall mainly be concerned with the realizations
 determined by $ f(B)=\sqrt{1-B},$ but shall consider other realizations
 as well. While the realization defined by $ f(B)=\sqrt{1-B}$ can be
 treated exactly, the other realizations are very difficuilt to treat
 in that manner, although some types of realizations admit exact treatment
  under some special cicumstances. For example, when the deformation
  of spacetime is of snyder-type, then it is possible to carry out an
  exact analysis not only for the function $ f(B)=\sqrt{1-B},$ but
   for the function $ f(B)=1,$ as well.
Due to technical difficuilties related to exact treatment of the
general type of realization, the
final part of the paper will be devoted to perturbative treatment
of that case, which is determined by the form of function $f$ that is  consistent
with all imposed requirements, but otherwise arbitrary.
This perturbative treatment of the general case of realization will result
with all necessary Hopf-algebraic ingredients, determined up to second order in the deformation parameters $a$ and $s.$


\section{Enveloping algebra modules}

In this section we investigate modifications which affect the algebra of
functions upon deforming the spacetime structure. Since the algebra (\ref{2.1})
mixes spacetime coordinates with Lorentz generators, the algebraic structure
obtained upon deformation is much richer than the original one. As a
consequence, these two structures will not be isomorphic to each
other. This is somewhat unpleasant situation because it does not allow
us to make full correspondence between commutative and noncommutative
algebras, the feature which is anyway common for Moyal and pure
$\kappa$-deformation $(s = 0)$. It would be thus convenient to find a
way out of this situation and to look for the means by which the isomorphism could be again established.
Before being able to do that, we have to recapitulate some basic notions 
from previous sections and to introduce few new ones.
Firstly, it is assumed in our considerations that noncommutative functions are
infinitely derivable, i.e. that they belong to the set $C^{\infty}
({\mathbb{R}}^{1,3})$ of infinitely derivable functions defined  on Minkowskian spacetime manifold.
As already pointed in previous section, they can be arranged into an
algebra, whose multiplication operation is given by the
standard multiplication of functions, to form
an algebra of functions 
in commutative coordinates, $\A_X$. This algebra can also be considered  as
the commutative enveloping algebra in $X_\mu$.
In the similar way, as already defined in section 2,
 with $\hat{\A}_{\x}$ we denote the noncommutative enveloping algebra in
$\x_\mu$, which is set to represent the
algebra of functions in noncommutative coordinates.
Besides these two kinds of algebras, $\A_X$ and $\hat{\A}_{\hat{x}}$,
there is still a third type of algebra that is relevant in our
discussions and that one is denoted by ${\A}_{X,\star}$ and
represents a noncommutative algebra of
functions in commutative coordinates, but with the $\star$-product
having the role of a multiplication operation.
We shall return to the algebra ${\A}_{X,\star}$ and definition of the
star product shortly,
after we introduce all other necessary conceptual pieces in the scheme we work in.

Let us further denote by $H \equiv H(X_{\mu},D_\mu)$ an undeformed
Heisenberg algebra generated by $X_\mu, D_\mu$ and defined by
equations (\ref{2.5}),(\ref{2.9}),(\ref{2.10}). The corresponding universal enveloping algebra
can then be denoted by $\U(H).$ 

Let us next introduce the unit element $1 \in {\A}_{X} $ and define the action of
Poincar\'{e} generators $D_\mu$ and $M_{\mu \nu}$ on $1$ as
\begin{equation} \label{actionde} 
D_{\mu} \triangleright 1 = 0, \qquad  M_{\mu\nu} \triangleright 1 = 0.
\end{equation}
Hence, the action of Poincar\'{e} algebra $\P,$ generated by
$D_\mu$ and $M_{\mu\nu},$ on $1$ is: $\P\triangleright 1 = 0$.
The unit element $1$ is here assumed to be a unit element in the
algebra $\A_X$, understood as a module for the algebra
$\U(H)~$ ($\A_X$ is $\U(H)$ module).

Bearing this in mind, we can define the action of $\A_X$ on $1$ as a Fock-like vector space
denoted by $\A_X \triangleright 1,$ with the property
\begin{equation}
 X_{\mu} \triangleright 1 = X_{\mu} \quad \mbox{and consequently}
 \quad  \phi(X) \triangleright 1 = \phi(X),
\end{equation}
satisfied for any $\phi(X) \in \A_X$. Hence $\A_X \triangleright 1 = \A_X.$
Similarly we define the action $\U(H) \triangleright 1$. Then the $\U(H) $ module is
defined by
\begin{equation}
\U(H) \triangleright \A_X ~ = ~ \U(H) \A_X \triangleright 1 ~ = ~ \A_X.
\end{equation}

By following the same steps,
 we introduce deformed Heisenberg algebra
$\hat{H}(\x_{\mu},D_\mu),$ defined by equations (\ref{2.1}),(\ref{2.5}),(\ref{2.6}). The
corresponding universal enveloping algebra can then be denoted with
${\U}(\hat{H})$.
As defined earlier, the noncommutative enveloping algebra in
$\x_\mu$ is denoted by $\hat{\A}_{\hat{x}}$.
The action of  $D_\mu$  and  $M_{\mu\nu}$ is given by  (\ref{actionde}),
so that we can also define the action of $\hat{\A}_{\x}$ on $1$ as a
Fock-like  vector space denoted by $\hat{\A}_{\x} \triangleright 1.$

In the previous section, we have seen that the commutation relations
(\ref{2.1}) admit a class of realizations of the form
\begin{equation}
\x_{\mu} =X^\alpha \Phi_{\alpha\mu}(D). \nonumber
\end{equation}
The commutation relations (\ref{2.1}),(\ref{2.6}) and the general form (\ref{2.7})
of realizations imply that we have
\begin{equation}
[\x_{\mu},\x_{\nu}]\triangleright 1=
i(a_{\mu}\x_{\nu}-a_{\nu}\x_{\mu})\triangleright 1,
\end{equation}
as well as
\begin{equation} \label{demiaction}
[D_{\mu},\x_{\nu}] \triangleright 1 = \eta_{\mu\nu} \triangleright 1.
\end{equation}
and
\begin{equation} \label{oneelementmonomial}
 \x_{\mu} \triangleright 1 = X_{\mu}.
\end{equation}
We point out that by shifting the derivative $D_{\mu}$ to the most
right in the element of ${\U}(\hat{H}),$ acting on $1,$ and by using 
(\ref{demiaction}) and Jacobi identities, we can establish the equivalence
\begin{equation}
  {\U}(\hat{H}) \triangleright 1 ~ \equiv ~ \hat{\A}_{\x} \triangleright 1.
\end{equation}

In the similar way as done with the monomial
(\ref{oneelementmonomial}), it is also possible to calculate the action on unit element $1$ by an
arbitrary monomial in noncommutative coordinates. All that is
necessary to do this is to replace NC coordinates according to (\ref{2.7})
  and then, by successively applying the commutation relation
 (\ref{2.10}), shift the derivative $D_{\mu}$ to the
most right, to get
\begin{equation} \label{multielementmonomial}
 \x_{\mu_{\Pi(1)}}...\x_{\mu_{\Pi(m)}}\triangleright 1=
 X_{\mu_{1}}...X_{\mu_{m}}+P_{(m-1),\Pi}(X)
\end{equation}
where $P_{(m-1),\Pi}(X)$ is a polynomial in $X_{\mu}$ of order
$(m-1)$, $ m\in N $ and $\Pi$ is permutation, $\Pi\in S_m$.

It is obvious from  (\ref{multielementmonomial}) that the space $\hat{\A}_{\x} \triangleright 1$ is
  contained within $\A_X$.
  However, to show that opposite is also true, i.e. that $\A_X$ is
  contained within $\hat{\A}_{\x} \triangleright 1,$  we have to invoke the
  inverse of the  transformation (\ref{2.7}). It is assumed that the
  inverse of (\ref{2.7}) exists, so that it can generally be written in the form
\begin{equation} \label{generalinverse}
X_{\mu} =\x^\alpha (\Phi^{-1})_{\alpha\mu}(D).
\end{equation}
For example, for the case of realization (\ref{3.2}), the
corresponding inverse will be given in the next section (see Eq.(\ref{3.12})).
Now, by taking the arbitrary element of $\A_X,$ replacing all
commutative coordinates according to prescription
(\ref{generalinverse}) and by shifting all derivatives to the most
right, we regularly finish with some element in $\hat{\A}_{\x} \triangleright 1,$
showing that $\A_X$ is indeed contained within $\hat{\A}_{\x} \triangleright 1.$
Thus, as the conclusion, we have that
the space $\hat{\A}_{\x}\triangleright 1$ is isomorphic to
$\A_{X},$ as a vector space.

Hence, the space $\hat{\A}_{\x}$ is larger than space
$\hat{\A}_{\x}\triangleright 1$ and the space
$\hat{\A}_{\x} \triangleright 1$ is isomorphic to $\A_{X}$ as a
vector space.
 (Note that $\hat{\phi}(\x)\triangleright 1$ can be
identified with
 $\hat{\phi}(\x)|0>$ in analogy with Fock-like space).
Overall consistency follows from the Jacobi identities and, in addition,
the corresponding ${\U}(\hat{H})$ module can then be defined by:
\begin{equation}
  U(\hat{H}) \triangleright (\hat{\A}_{\x} \triangleright
 1) = (U(\hat{H}) \hat{\A}_{\x}) \triangleright
 1 = \hat{\A}_{\x} \triangleright 1
\end{equation}
and generally we can write
\begin{equation} \label{actionphi} 
\hat{\phi}(\x) \triangleright 1 = \phi(X),
\end{equation}
for any  $ \hat{\phi}(\x) \in \hat{\A}_{\x}$ and for ${\x}$ given by (\ref{2.7}). 

Let us now turn to definition of the star product, as promised before.
Thus, if we have two associations of the form (\ref{actionphi}),
namely, $\hat{\phi}(\x)\triangleright 1 = \phi(X)$ and
$\hat{\psi}(\x)\triangleright 1=\psi (X),$ for two noncommutative functions
$\hat{\phi}(\x)$ and $\hat{\psi}(\x),$  then the star product is
defined by
\begin{eqnarray} \label{starproductdefinition}
\hat{\phi}(\x)\hat{\psi}(\x) \triangleright
1 ~ & = & ~ \hat{\phi}(\x)\triangleright(\hat{\psi}(\x)\triangleright 1) \nonumber \\
 ~& =& ~ \hat{\phi}(\x)\triangleright \psi(X) 
 ~ = ~  \phi (X) \star \psi(X),
\end{eqnarray}
where it is understood that ${\x}$ is given by (\ref{2.7}).
The vector space $\A_X$ together with the star product (\ref{starproductdefinition})
constitutes a noncommutative algebra, which we denote by ${\A}_{X,\star}$.
The algebra ${\A}_{X,\star}$ is identical to noncommutative algebra
$\hat{\A}_{\x} \triangleright 1$ 
and is generally nonassociative, which can be seen
by generalizing the definition (\ref{starproductdefinition}) of the star product
to a star product of three fields. In this case we come up with
\begin{eqnarray} \nonumber
\hat{\phi_1}(\x)\hat{\phi_2}(\x)\hat{\phi_3}(\x)\triangleright
1=(\hat{\phi_1}(\x)\hat{\phi_2}(\x))\triangleright
(\hat{\phi_3}(\x)\triangleright 1)=
\hat{\phi_1}(\x)\triangleright(\hat{\phi_2}(\x)\triangleright(\hat{\phi_3}(\x)\triangleright
1))\\= \nonumber \hat{\phi_1}(\x)\triangleright(\phi_2(X)\star
\phi_3(X))=\hat{\phi_1}(\x)\hat{\phi_2}(\x) \triangleright
  \phi_3(X) = \phi_1(X)\star(\phi_2(X)\star\phi_3(X)) 
\end{eqnarray}
and specifically, when looking a product of NC coordinates, we have
\begin{equation}
 \x_1 \x_2 \x_3 \triangleright
 1 = \x_1 \triangleright (\x_2 \triangleright ( \x_3 \triangleright 1))
 = X_1 \star (X_2 \star X_3).
\end{equation}
showing that if $s\neq0,$ the star product is generally
non-associative \cite{Girelli:2009ii},\cite{Battisti:2010sr},\cite{Girelli:2010wi},\cite{girelli}.

In addition, it should be noted that translation generators $D_{\mu}$ act on the elements of
$\hat{\A}_{\x}$ as
\begin{equation}
D_{\mu}\hat{\phi}(\x) \triangleright 1 = 
D_{\mu} \triangleright (\hat{\phi}(\x) \triangleright1 )
 = D_{\mu} \triangleright \phi(X) = \frac{\partial \phi(X) }{\partial  X^{\mu}} \nonumber
\end{equation}
and
\begin{equation}
D_{\mu}\hat{\phi}(\x)\hat{\psi}(\x) \triangleright
1 = D_{\mu} \triangleright
(\hat{\phi}(\x)\hat{\psi}(\x)\triangleright1) = D_{\mu} \triangleright
(\phi (X)\star \psi(X)) \nonumber
\end{equation}
and that the action of Lorentz generators $M_{\mu \nu}$ is defined by
\begin{eqnarray}
M_{\mu \nu}\hat{\phi}(\x) \triangleright 1 & =& 
M_{\mu \nu} \triangleright (\hat{\phi}(\x) \triangleright1 )
  = M_{\mu \nu} \triangleright \phi(X), \nonumber \\
M_{\mu \nu}\hat{\phi}(\x)\hat{\psi}(\x) \triangleright
1 & =& M_{\mu \nu} \triangleright
(\hat{\phi}(\x)\hat{\psi}(\x) \triangleright1 ) = M_{\mu \nu} \triangleright
(\phi (X)\star \psi(X)). \nonumber
\end{eqnarray}

Finally, we define another action $\circ$ on the same unit element 1, defined by
\begin{equation} \label{adefinition}
D_{\mu}\circ 1=0, \qquad  M_{\mu\nu}\circ 1=0
\end{equation}
and
\begin{equation} 
\x_{\mu}\circ 1= \x_{\mu},
\end{equation}
where the last relation provides a construction of Fock-like space $\hat{\A}_{\x}\circ 1$.
The action denoted by $\circ$  implicitly includes the information
that, besides definition (\ref{adefinition}),  one has to
use the inverse transformation (\ref{generalinverse}) in calculating the
expressions of the form $\Phi(X(\x,D))\circ 1,$ in the same way as the action $\triangleright$
implicitly includes the instruction of using direct transformation (\ref{2.7}),
when calculating the expressions such as $\hat{\Phi}(\x)\triangleright 1.$
 In this way, by using the inverse map (\ref{generalinverse}),
it is easy to show that
\begin{equation} \label{aa1a}
X_{\mu} \circ 1 = \x^\alpha (\Phi^{-1})_{\alpha\mu}(D) \circ 1 =
 \x^\alpha \circ ((\Phi^{-1})_{\alpha\mu}(D) \circ 1) =
\x_\mu \circ 1 = \x_\mu.
\end{equation}
If we assume that the relation
\begin{equation}
\hat{\Phi}(\x)\triangleright 1=\Phi(X) \nonumber
\end{equation}
is satisfied, then Eq.(\ref{aa1a})
 could possibly tempt one to conclude that
\begin{equation} \label{aa2a}
\Phi(X(\x,D))\circ 1=\hat{\Phi}(\x). 
\end{equation}
However, this is not generally true, since already for monomial
including the product of two coordinates, this relation fails to hold.
For example, in the general case when $s \neq 0,$ we have
\begin{eqnarray} \label{aa1aexample}
  X_{\mu} X_{\nu} \circ 1 & =& (\x^\alpha (\Phi^{-1})_{\alpha\mu}(D)) (\x^\beta (\Phi^{-1})_{\beta\nu}(D)) \circ 1  \nonumber \\
 &=& (\x^\alpha (\Phi^{-1})_{\alpha\mu}(D)) \x_\nu \circ 1 =  (\x_\mu \x_\nu + P_{\mu \nu} (\x) ) \circ 1,  \nonumber
\end{eqnarray}
where $P_{\mu \nu} (\x) $ is some  polynomial linear in $\x$.
It is only in the special case when $s = 0$ that the above 
relation takes the form (\ref{aa2a}). That this is so can be inferred by inserting
 the explicit form (\ref{3.12}) of the inverse
map (\ref{generalinverse}) into l.h.s. of the above expression and
proceed further according to definition (\ref{adefinition}).

Although relation (\ref{aa2a}) is not valid generally, the correct
 relationship between commutative and noncommutative functions can be stated in
the following way: If the function $\hat{\Phi}(\x) $ from
$\hat{\A}_{\x} $ satisfies the relation
\begin{equation} \label{aa3aprva}
\hat{\Phi}(\x)\triangleright 1=\Phi(X),
\end{equation}
for some $\Phi(X)$ in $\A_X,$ then use of the inverse map 
(\ref{generalinverse}) leads generally not to (\ref{aa2a}), but
instead leads to
\begin{equation} \label{aa3a}
\Phi(X(\x,D))\circ 1=\hat{\Phi}(\x) \circ 1,
\end{equation}
showing that generally $\hat{\Phi}(\x) \circ 1 \neq \hat{\Phi}(\x).$
An example that can serve to illustrate this last statement is provided by
analysing  the action
$$
 [\x_{\mu}, \x_{\nu}] \circ 1 = i(a_{\mu}\x_{\nu}-a_{\nu}\x_{\mu}).
$$
A simple analysis shows that relation $[\x_{\mu}, \x_{\nu}] \circ 1 = [\x_{\mu}, \x_{\nu}] $ 
 is fulfilled only for $s = 0,$ while
in the general case when $s \neq 0,$ it is not satisfied.

Described relationship thus establishes the isomorphism
between spaces $\hat{\A}_{\x}\circ 1$ and
$\hat{\A}_{\x}\triangleright 1$ at the level of vector space.
The space $\hat{\A}_{\x}\circ 1$ as a vector space is thus isomorphic to
$\hat{\A}_{\x}\triangleright1$, since
\begin{equation}
\x_{\mu_{\Pi(1)}}...\x_{\mu_{\Pi(m)}}\circ 1=
\x_{\mu_{1}}...\x_{\mu_{m}}\circ 1+\hat{P}_{(m-1),\Pi}(\x) \circ 1,
\end{equation}
where $\hat{P}_{(m-1),\Pi}(\x)$ is a polynomial in $\x_{\mu}$ of
order $(m-1)$, $ m\in N $ and $\Pi$ is permutation, $\Pi\in S_m$.
Hence, the algebra $\hat{\A}_{\x}$ is larger than space
$\hat{\A}_{\x}\circ 1$ and the space $\hat{\A}_{\x}\circ 1$ is
isomorphic to $\A_{X}$ as a vector space.


We briefly summarise the basic results of this section.
We have shown that although the vector spaces $\A_{X}$ and $\hat{\A}_{\x}$ of commutative and noncommutative functions
are not isomorphic to each other, it is possible to find
  vector spaces that are isomorphic to the space $\A_{X}$. 
 These are given by the actions $\hat{\A}_{\x} \triangleright 1 $
 and $\hat{\A}_{\x} \circ 1$ of the algebra $\hat{\A}_{\x}$ to the
 unit element $1.$  We emphasize that 
  given a specific realization $\Phi_{\alpha\mu}(D)$, Eq.(\ref{2.7}), of noncommutative
 coordinates, the necessary and sufficient
condition to have the above isomorphism realised at the level of vector spaces
 is the existance of the inverse map $(\Phi^{-1})_{\alpha\mu}(D)$
in (\ref{generalinverse}).
Since the realizations (\ref{3.2}) that are of interest to us in this paper have a well defined
 inverse map 
(see Eq.(\ref{3.12}) below), we can regularly establish the
isomorphism in our case. This isomorphism is compactly described by the
two mutually entangled relations (\ref{aa3aprva}) and (\ref{aa3a}).

As an example, instead of  (\ref{2.6}), the most general relations
including coordinates and derivatives  
can be written in the form
\begin{equation}
 {\hat{\partial}}_i {\x}_j - q_{ij} {\x}_i {\hat{\partial}}_i = \Phi_{ij}(\hat{\partial}),
\end{equation}
describing the $q$-deformation of the Heisenberg algebra.
For $\Phi_{ij} = {\delta}_{ij}, $ these relations were classified in \cite{fronsdal}.
In this case, the choice with $q_{ij} = \pm 1$ leads to Heisenberg or
Clifford algebra, or in other words to Bose or Fermi statistics,
respectively. 
The situation with generic $q_{ij}$ leads to infinite statistics
 algebras \cite{greenberg}.
In this example there is no mapping at all between $\A_{X}$ and $\hat{\A}_{\x}$
and noncommutative coordinates and derivatives cannot be expressed in
terms of commutative ones, showing that the $q$-deformed Heisenberg
algebra is much larger than the undeformed Heisenberg algebra.
 Unlike the situation we have in
this paper, in the case with generic $q_{ij}$ the isomorphism cannot be established even between 
$\hat{\A}_{\x} \circ 1 $ and $\A_{X}$.



\section{Invariants under Lorentz algebra}

As in the ordinary commutative Minkowski space, here we can also take
the operator $P^2 = P_\alpha P^{\alpha} = -D^2$ as a Casimir operator,
playing the role of an invariant in
noncommutative Minkowski space. In doing this, we introduced the
momentum operator  $ P_\mu = -iD_\mu.$ In this case, arbitrary
function $F(P^2)$ of Casimir also plays the role of invariant, namely
$[M_{\mu\nu},F(P^2)]=0.$
 However, unlike the ordinary
Minkowski space, in NC case we have a freedom to introduce still
another invariant by generalizing the standard notion of d'Alambertian operator
to the generalized one required to satisfy
\begin{align} \label{3.5}
[M_{\mu\nu}, \square]&=0,  \\
[\square, \x_{\mu}]&= 2D_{\mu}.
\end{align}
The general form of the generalized d'Alambertian operator $\square,$
valid for the large class of realizations (\ref{3.2}), which are characterized
by an arbitrary function $f(B),$ can be written in a compact form as
\begin{equation} \label{3.6}
  \square = \frac{1}{a^2 - s} \int_0^B \frac{dt}{f(t)-t\gamma_2 (t)},
\end{equation}
where $\gamma_2 (t)$ is given in (\ref{3.3}).
Due to the presence of the Lorentz invariance in NC Minkowski space
(\ref{2.1}), the basic dispersion relation is undeformed, i.e. it reads
$P^2 + m^2 =0$  for all $f(B).$
Specially, for $f(t)= \sqrt{1-t},$ we have $\gamma_2 (t)=0$ and,
consequently, the generalized d'Alambertian is given by
\begin{equation} \label{3.7}
  \square = \frac{2(1- \sqrt{1-B})}{a^2 - s}.
\end{equation}
It is easy to check that in the limit $a,s \rightarrow 0,$ we have the
standard result, $\square \rightarrow D^2,$ valid in undeformed
Minkowski space.

Lorentz symmetry provides us with the possibility of constructing the invariants.
In most general situation, for a given realization $\Phi_{\mu\nu},$
Eq.(\ref{3.2}), Lorentz invariants can as well be constructed out
of NC coordinates $\x_{\mu}.$

To proceed further with the construction of invariants in NC coordinates for a
 given realization $\Phi_{\mu\nu},$ it is also of interest to
 wright down the inverse of realization
(\ref{3.2}), namely,
\begin{equation} \label{3.12}
X_{\mu}
=[\x_\mu-i(a\x)\frac{1}{f(B)-B\gamma_2}D_\mu+(a^2-s)(\x D)
\frac{1}{f(B)-B\gamma_2}D_\mu\gamma_2]\frac{1}{-A+f(B)}.
\end{equation}
Since we know how to construct invariants out of commutative
coordinates and derivatives, namely, $X_{\mu}$ and $D_{\mu},$
relation (\ref{3.12}) ensures that the similar construction can be
carried out in terms of NC coordinates $\x_{\mu}.$ The same
construction also applies to tensors. All that is required is that
the general invariants and tensors, expressed in terms of
$X_{\mu}$ and $D_{\mu},$ have to be transformed into corresponding
invariants and tensors in NC coordinates $\x_{\mu}$ and $D_{\mu}$
with the help of the inverse transformation (\ref{3.12}), which,
in accordance with Eq.(\ref{2.7}), can compactly be written as
$X_{\mu}= \x_{\alpha} {(\Phi^{-1})}_{\alpha\mu}.$ General tensors
in NC coordinates can now be built from tensors $X_{\mu_1}\cdot
\cdot \cdot X_{\mu_k}D_{\nu_1}\cdot \cdot \cdot D_{\nu_l}$ in
commutative coordinates by making use of the inverse
transformation (\ref{3.12}), $X_{\mu_1}\cdot \cdot \cdot
X_{\mu_k}D_{\nu_1}\cdot \cdot \cdot D_{\nu_l} = \x_{\beta_1}
{(\Phi^{-1})}^{\beta_1}_{~~\mu_1}\cdot \cdot \cdot \x_{\beta_k}
{(\Phi^{-1})}^{\beta_k}_{~~\mu_k}D_{\nu_1}\cdot \cdot \cdot
D_{\nu_l}. $ The same holds for the invariants. For example,
following the described pattern, we can construct the second order
invariant in NC coordinates in a following way. Knowing that the
object $X^2 = X_{\alpha}X^{\alpha}$ is a Lorentz second order
invariant, $[M_{\mu \nu}, X_{\alpha}X^{\alpha}]=0,$ the
cooresponding second order invariant $\hat{I}_2$ in NC coordinates
can be introduced as $\hat{I}_2 = X_{\alpha}X^{\alpha}\circ 1. $
After use has been made of (\ref{3.12}), simple calculation gives
$\hat{I}_2$ expressed in terms of NC coordinates, $\hat{I}_2 =
\x_{\alpha}\x^{\alpha}\circ 1-i(n-1) a_{\alpha}\x^{\alpha}\circ
1.$ It is now easy to check that the action of Lorentz generators
on $\hat{I}_2$ gives $M_{\mu \nu} \circ \hat{I}_2 =0,$ confirming the
validity of the construction.\\

{\bf Remark}\\

It is important to realize that NC space with the type of
noncommutativity (\ref{2.1}) can be mapped to Snyder space with the
help of transformation
\begin{equation} \label{3.13}
   \hat{\tilde{x}}_{\mu} = \hat{x}_{\mu} - i a^{\alpha}M_{\alpha\mu},
\end{equation}
generalizing the transformation used in \cite{KowalskiGlikman:2002jr}
to map $\kappa$-deformed space to Snyder space.
After applying this transformation, we get
\begin{equation} \label{3.14}
   [\hat{\tilde{x}}_{\mu}, \hat{\tilde{x}}_{\nu}] = (s-a^2) M_{\mu\nu},
\end{equation}
\begin{equation} \label{3.15}
   [M_{\mu\nu}, \hat{\tilde{x}}_{\lambda}] = \eta_{\nu\lambda}\hat{\tilde{x}}_{\mu}
               - \eta_{\mu\lambda}\hat{\tilde{x}}_{\nu}.
\end{equation}
The mapping (\ref{3.13}) between spaces with symplectic structures of
$\kappa$-Snyder and Snyder type, respectively, has properties that obviously depend
on the mutual relations between the deformation parameters. This can best
 be seen when the question of isomorphism is raised. Hence, for $s = 0, $
 the above spaces are not isomorphic, while for $s \neq 0, $
with an additional condition that $a^{2} \neq s, $ these spaces are isomorphic.
Finally, for $a^{2} = s $ there is no isomorphism and this situation
represents a singular point, with an effective Snyder deformation parameter 
 equal to zero, i.e. leading to commutative geometry.

The Lorentz generators are expressed in terms of this new coordinates as
\begin{equation} \label{3.16}
M_{\mu\nu}=(\hat{\tilde{x}}_{\mu}D_{\nu}- \hat{\tilde{x}}_{\nu}D_{\mu})\frac{1}{f(B)},
\end{equation}
and $\hat{\tilde{x}}_{\mu}$ alone, allows the representation
\begin{equation} \label{3.17}
 \hat{\tilde{x}}_{\mu}=X_{\mu} f(B) -(a^2-s)(XD)D_\mu\gamma_2,
\end{equation}
in accordance with (\ref{3.2}). The results, starting with the
mapping (\ref{3.13}) and all down through Eq.(\ref{3.17}), are
valid not only for the particularly interesting choice $f(B)=\sqrt{1-B},$ but instead are
valid for an arbitrary function satisfying the boundary condition
$f(0)=1.$

\section{Leibniz rule and Hopf algebra}

  The symmetry
 underlying deformed Minkowski space, characterized by the commutation relations
   (\ref{2.1}), is the deformed Poincar\'{e} symmetry which can most
 conveniently be described in terms of quantum Hopf algebra.
 As was seen in relations (\ref{2.2a}),(\ref{2.5}) and (\ref{2.5a}),
 the algebraic sector of this deformed symmetry is the same as that of undeformed
Poincar\'{e} algebra. However, the action of Poincar\'{e} generators
 on the deformed Minkowski space is deformed,
  so that the whole deformation
 is contained in the coalgebraic sector. This means that the Leibniz
 rules, which describe the action of $M_{\mu\nu}$ and $D_{\mu}$ generators,
will no more have the standard form, but instead will be deformed
and will depend on a given $\Phi_{\mu\nu}$ realization.

  Generally we find that in a given $\Phi_{\mu\nu}$ realization we can write
\begin{equation} \label{4.1}
 e^{ik\x}\triangleright 1 = ~ e^{iK_{\mu}(k)X^{\mu}}
\end{equation}
and
\begin{equation} \label{4.2}
 e^{ik\x} \triangleright e^{iqX} ~ = ~ e^{iP_{\mu}(k,q)X^{\mu}},
\end{equation}
where the action on the unit element is defined in section 4, Eqs.(\ref{actionde}),(\ref{actionphi}) and
 $k\x = k^{\alpha}X^{\beta}\Phi_{\beta\alpha} (D). $
The quantities $K_{\mu}(k)$ are readily identified as $K_{\mu}(k) = P_{\mu}(k,0)$
and $P_{\mu}(k,q)$ can be found by calculating the expression
\begin{equation} \label{4.3}
  P_{\mu}(k,-iD) ~ = ~ e^{-ik\x} (-iD_{\mu}) e^{ik\x},
\end{equation}
where it is assumed that at the end of calculation the identification
$q=-iD$ has to be made.
One way to explicitly evaluate the above expression is by using the
standard {\it{ad}}-expansion perturbatively, order by order. To avoid this tedious
procedure, we can turn to much more elegant method to obtain the
quantity $P_{\mu}(k,-iD)$. This consists in writing the differential equation
\begin{equation} \label{4.4}
 \frac{dP_{\mu}^{(t)}(k,-iD)}{dt} ~ = ~
 \Phi_{\mu\alpha}(iP^{(t)}(k,-iD)) k^{\alpha},
\end{equation}
satisfied by the family of operators $P_{\mu}^{(t)}(k,-iD),$ defined as
\begin{equation} \label{4.5}
  P_{\mu}^{(t)}(k,-iD) ~ = ~ e^{-itk\x} (-iD_{\mu}) e^{itk\x}, \qquad 0\leq t\leq 1,
\end{equation}
and parametrized with the free parameter $t$ which belongs to the interval
$0\leq t\leq 1.$ The family of operators (\ref{4.5}) represents the
generalization of the quantity $P_{\mu}(k,-iD),$ determined by (\ref{4.3}),
namely, $P_{\mu}(k,-iD) = P_{\mu}^{(1)}(k,-iD). $ Note also that
solutions to
differential equation (\ref{4.4}) have to satisfy the boundary
condition $P_{\mu}^{(0)}(k,-iD) = -iD_{\mu} \equiv q_{\mu}.$
 The function $\Phi_{\mu\alpha}(D)$ in (\ref{4.4}) is deduced from (\ref{3.2}) and
 reads as
\begin{equation} \label{4.6}
  \Phi_{\mu\alpha}(D) =\eta_{\mu\alpha}(-A+f(B))+ia_{\mu}D_{\alpha}-(a^2-s)D_{\mu}D_\alpha\gamma_2.
\end{equation}
All results so far are written for the most general
type of realizations. A complete solution requires integration of Eq.(\ref{4.4}),
which may not generally be so easy problem to handle and cannot be
solved exactly for an arbitrary admissible function $f(B).$ We shall
take care of this most general case in the final part of the paper, where the
perturbative solution to Eq.(\ref{4.4}) will be found, valid through
the second order in deformation parameters. There are however few
exceptional choices for the function $f(B)$ that allow for an exact solution. One
example of such case is the function $f(B) = \sqrt{1-B}, $ which
anyway appers frequently in the literature. For this particular case,
that will below be solved exactly, we have $\gamma_2 = 0$ and consequently
Eq.(\ref{4.4}) reads
\begin{equation} \label{4.7}
 \frac{dP_{\mu}^{(t)}}{dt} ~ = ~ k_{\mu} \bigg[ aP^{(t)} +
 \sqrt{1+ (a^2 -s){(P^{(t)})}^2} ~ \bigg] -a_{\mu} kP^{(t)},
\end{equation}
where we have used an abbreviation $P_{\mu}^{(t)} \equiv P_{\mu}^{(t)}(k,-iD).$
The exact solution to differential equation (\ref{4.7}), which obeys the
required boundary conditions, looks as
\begin{eqnarray} \label{4.8}
 P_{\mu}^{(t)}(k,q) & = & q_{\mu} + \left( k_{\mu} Z^{-1}(q) -a_{\mu}
 (kq)  \right) \frac{\sinh(tW)}{W}  \\
        &+& \bigg[ \left(k_{\mu}(ak) - a_{\mu}k^2 \right) Z^{-1}(q)
     + a_{\mu}(ak) (kq) - sk_{\mu} (kq) \bigg] \frac{\cosh (tW)-1}{W^2}. \nonumber
\end{eqnarray}
In the above expression we have introduced the following abbreviations,
\begin{eqnarray} \label{4.9}
 W &=& \sqrt{{(ak)}^2 - s k^2}, \\
  Z^{-1} (q) & =& (aq) + \sqrt{1 + (a^2 - s)q^2}
\end{eqnarray}
and it is understood that quantities like $(kq)$ mean the scalar
product in a Minkowski space with signature
 $~ \eta_{\mu\nu} = diag (-1,1,\cdot \cdot \cdot, 1)$.
Now that we have $P_{\mu}^{(t)}(k,q),$ the required quantity
$P_{\mu}(k,q)$ simply follows by setting $~t=1~$
and finaly we also get
\begin{eqnarray} \label{4.10}
  K_{\mu} (k)  =
 \bigg[ k_{\mu} (ak)   -
  a_{\mu} k^2 \bigg] \frac{\cosh W -1}{W^2}
    + k_{\mu} \frac{\sinh W}{W}.
\end{eqnarray}

We can now write the star product between arbitrary two plane waves in the
algebra ${\A}_{X,\star}$ as follows,
\begin{equation} \label{4.11}
  e^{ikX}~ \star ~ e^{iqX} ~ \equiv ~ e^{i K^{-1}(k) \x} \triangleright e^{iqX}
 ~ = ~ e^{i {{\mathcal{D}}_{\mu} (k,q)}X^{\mu}},
\end{equation}
where
\begin{equation} \label{4.12}
   {\mathcal{D}}_{\mu} (k,q) ~ = ~ P_{\mu} (K^{-1}(k),q),
\end{equation}
with $K^{-1}(k)$ being the inverse of the transformation (\ref{4.10}).
In writing the star product (\ref{4.11}), we have applied
Eq.(\ref{4.1}) and the definition (\ref{starproductdefinition}) of
the star product.
It is further possible to show that
quantities $~Z^{-1}(k)~$ and $~\square(k) ~$ can be expressed in terms
of quantity $K^{-1}(k)$ as
\begin{equation} \label{4.13}
  Z^{-1} (k)  \equiv (ak) + \sqrt{1 + (a^2 - s)k^2} = \cosh W( K^{-1}(k))
       + a K^{-1}(k) \frac{\sinh W( K^{-1}(k))}{W( K^{-1}(k))},
\end{equation}
\begin{equation} \label{4.14}
   \square(k) \equiv \frac{2}{a^2-s}\bigg[ 1- \sqrt{1 + (a^2 - s)k^2}
   \bigg] = 2 {(K^{-1}(k))}^2 \frac{1-
 \cosh W( K^{-1}(k)) }{W^2( K^{-1}(k))},
\end{equation}
where $~W( K^{-1}(k))  ~$ is given by (\ref{4.9}), or explicitly
\begin{equation} \label{4.15}
  W( K^{-1}(k)) = \sqrt{{(aK^{-1}(k))}^2 -s{(K^{-1}(k))}^2}.
\end{equation}

  With the exact solutions (\ref{4.8}) and (\ref{4.10}), corresponding
   to realization $f(B) = \sqrt{1-B}, $ it is possible to determine
   all the ingredients that define Hopf algebra in the case of that
   particular realization. As a first step,
   the function $~{\mathcal{D}}_{\mu} (k,q)~$ determines \cite{Meljanac:2009fy} a deformed
   Leibniz rule and the corresponding coproduct $\triangle D_{\mu}$ in
   the following way,
\begin{equation} \label{coproductde}
  \triangle D_{\mu} = i{\mathcal{D}}_{\mu} (-iD \otimes 1,1 \otimes (-iD) ).
\end{equation}
Relations (\ref{4.13}) and (\ref{4.14}) are useful in obtaining the
   expression for the coproduct. However, in the general case of
   deformation, when both parameters $a_{\mu}$ and $s$ are different
   from zero, it is quite a difficuilt task to obtain a closed form
   for $\triangle D_{\mu},$ so we give it in a form of a series
   expansion up to second order in the deformation parameter $a,$
\begin{eqnarray} \label{4.16}
  \triangle D_{\mu} &=&  D_{\mu}\otimes \mathbf{1} +\mathbf{1}\otimes
 D_{\mu} \nonumber \\
 & -& iD_{\mu} \otimes aD +
 ia_{\mu} D_{\alpha} \otimes D^{\alpha}
 -\frac{1}{2}(a^2-s)D_{\mu}\otimes D^2 \\
 & -& a_{\mu} (aD)D_{\alpha} \otimes D^{\alpha}
   +\frac{1}{2} a_{\mu} D^2 \otimes aD
   +\frac{1}{2} s D_{\mu} D_{\alpha} \otimes D^{\alpha} + {\mathcal{O}}(a^3). \nonumber
\end{eqnarray}
  Since in the case of a pure Snyder deformation ($a = 0$) the coproduct for Lorentz generators is undeformed,
in the case of general deformation ($a,s \neq 0$), the same coproduct
will be identical as in the case of pure $\kappa$-deformation,
\begin{eqnarray} \label{coproductmo}
\triangle M_{\mu\nu} &=& M_{\mu\nu}\otimes
\mathbf{1}+\mathbf{1}\otimes M_{\mu\nu} \nonumber \\
&+& ia_{\mu}\left(D^{\lambda}-\frac{ia^{\lambda}}{2}\square\right)\,
Z\otimes
M_{\lambda\nu}-ia_{\nu}\left(D^{\lambda}-\frac{ia^{\lambda}}{2}\square\right)\,
Z\otimes M_{\lambda\mu}. 
\end{eqnarray}
 As we
shall see at the end of the paper, this result that is above
established in the special case of realization when  $f(B) =
\sqrt{1-B} $ and which says that the coproduct for
Lorentz generators in the case of general deformation ($a,s \neq 0$)
is the same as the coproduct for Lorentz generators in the case of 
$\kappa$-deformation is, in fact, a most
general result, valid for all  realizations, i.e. for all functions
$f(B),$ consistent with imposed requirements, because, for Snyder
deformation, the coproduct 
 for Lorentz generators is undeformed \cite{Battisti:2010sr},
no matter which realization is used.
The same type of reasoning applies when one is concerned with
obtaining the antipodes for Lorentz generators \cite{Meljanac:2010ps}. Namely, since it is
known \cite{Battisti:2010sr} that the antipodes for Lorentz generators
are also undeformed in the case of pure Snyder deformation ($a = 0$), the
same antipodes in the case of general, i.e. $\kappa $-Snyder
deformation ($a,s \neq 0$), will be identical to the antipodes for Lorentz
generators in a pure $\kappa $-deformation. This statement, like the
previous one, is true not only for the realization $f(B) = \sqrt{1-B},
$ but for all realizations satisfying the required conditions.
As far as the antipodes for translation generators are concerned, the function  $~{\mathcal{D}}_{\mu} (k,q)~$ 
also plays a crucial role, since these antipodes can immediately be obtained by solving the conditions,
\begin{equation} \label{antipode}
  {\mathcal{D}}_{\mu} (S(k),k ) = {\mathcal{D}}_{\mu} (k,S(k) ) = 0.
\end{equation}
On the other hand, the counits are all trivial.

Now that we have a coproduct, it is a straightforward procedure
 \cite{Meljanac:2006ui},\cite{Meljanac:2007xb}
to construct a star product between arbitrary
two functions $f$ and $g$ of commuting coordinates, generalizing in this way relation (\ref{4.11}) that
holds for plane waves. Thus, the general result for the star product, valid for the
NC space (\ref{2.1}), has the form
\begin{equation} \label{4.17}
 (f~ \star ~ g)(X) = \lim_{\substack{Y \rightarrow X  \\ Z \rightarrow X }}
 e^{X_{\alpha} [ i{\mathcal{D}}^{\alpha}(-iD_{Y},
 -iD_{Z}) - D_{Y}^{\alpha} -D_{Z}^{\alpha} ]} f(Y)g(Z).
\end{equation}
Although star product is a binary operation acting on the algebra of functions
defined on the ordinary commutative space, it encodes features
that reflect noncommutative nature of space (\ref{2.1}).

Following the line set up in section 4, it is worth noting that
relation (\ref{4.1})  gives a suitable example of the vector space-level isomorphism established
in section 4 between the algebras $\A_{X}$ and $\hat{\A}_{\x} \triangleright1$.
 Particularly, from (\ref{4.1}) it follows that
\begin{equation} \label{4.18a}
   e^{iK_{\mu}(k)X^{\mu}} \circ 1 ~ = ~ e^{ik\x} \circ 1,
\end{equation}
being in accordance with relations (\ref{aa3aprva}) and (\ref{aa3a}),
that is two relations that form the basis of described vector space-level isomorphism.
As already established before, for $s \neq 0$ the algebra
$\hat{\A}_{\x} \triangleright 1$
 is nonassociative. In some considerations that deal with
 noncommutative Snyder space ($a = 0$), which is also characterized by
 the nonassociative star product, there were attempts to enlarge 
 the original noncommutative spacetime by including Lorentz generators
 as well, with the aim of getting an associative star product
\cite{Girelli:2010wi},\cite{girelli}.
In the described setting the coordinates $ M_{\mu \nu}$ are
interpreted as coordinates describing extra dimensions and the Snyder
space with nonassociative algebra structure is considered as a
subspace of a bigger noncommutative space, which is generated by the coordinates
$({\x}_{\mu}, M_{\nu \lambda})$ and admits associative star product
\cite{Girelli:2010wi},\cite{girelli}.
In the following section we shall specialize the general results
obtained so far to four particularly interesting special cases.

\section{Special cases}

\subsection{1. case $(s =a^2)$}

 In this case, NC commutation relations take on the form
\begin{equation}
[\x_{\mu},\x_{\nu}]=i(a_{\mu}\x_{\nu}-a_{\nu}\x_{\mu})+a^2 M_{\mu
\nu}.
\end{equation}
Since we now have $f(B)=f(0)=1,$ the generalized shift operator becomes $Z^{-1} = 1-A$ and
the realizations (\ref{3.2}) and (\ref{2.17}) for NC coordinates and
Lorentz generators, respectively,
take on a simpler form, namely,
\begin{equation}
\x_{\mu}=X_{\mu}(1-A)+i(aX)D_{\mu},
\end{equation}
\begin{equation}
M_{\mu \nu}= (\x_{\mu}D_{\nu}-\x_{\nu}D_{\mu}) \frac{1}{1-A}.
\end{equation}
In addition, the generalized d'Alambertian operator
becomes a standard one, $~\square = D^2, ~$ and
 deformed Heisenberg-Weyl algebra (\ref{3.4}) reduces to
\begin{align}
[D_\mu, \hat{x}_\nu]=\eta_{\mu\nu}(1-A)+ia_\mu D_\nu.
\end{align}
Relations (\ref{2.13}) and (\ref{2.15}), that include generalized shift operator, also change in an
appropriate way. Particularly, we have
\begin{align}
[1-A,\hat{x}_\mu]=-ia_\mu (1-A)+a^2 D_\mu.
\end{align}
 We see from Eq.(\ref{4.16}) that the coproduct
for this case also simplifies since the term with $(a^2-s)$ drops out.

\subsection{2. case $(a=0)$}

 When $a^2=0,$ we have a Snyder type of noncommutativity,
\begin{equation}
[\x_{\mu},\x_{\nu}]=s M_{\mu \nu}.
\end{equation}
In this situation, our realization (\ref{3.2}) reduces precisely to
that obtained in \cite{Battisti:2008xy},\cite{Battisti:2010sr}. For a special choice when
 $f(B) =1,$ we have the realization
\begin{equation} \label{snyderrealizationtutu}
\x_{\mu}=X_{\mu}-s(XD)D_\mu,
\end{equation}
which is the  case that was also considered in \cite{Licht:2005rm}.
The solution to Eq.(\ref{4.4}) for $f(B)  =1 $ and $ a=0$
leads to the following coproduct for translation generators,
\begin{eqnarray} \label{snydercoproduct}
\Delta D_\mu  & =  & \frac{1}{1+sD_\nu\otimes D^\nu} \bigg( D_\mu\otimes1  \nonumber \\
  & +  & \frac{s}{1+\sqrt{1+sD^2 }}\,D_\mu D_\nu\otimes D^\nu+\sqrt{1+ sD^2}\otimes D_\mu \bigg).
\end{eqnarray}

In other interesting situation, when $f(B)=\sqrt{1-B},$ the general result
(\ref{3.2}) reduces to
\begin{equation} \label{maggiorerealizationtutu}
\x_{\mu}=X_{\mu} \sqrt{1+sD^2}.
\end{equation}
This choice of $f(B)$ is the one for which most of our results, through
all over
the paper, are obtained and which is one of the main objects of our
investigations. It is also considered by Maggiore \cite{Maggiore:1993rv}.
 For this case when $f(B)=\sqrt{1-B},$ the exact
result for the coproduct (\ref{4.12}) can be obtained and it is given by
\begin{equation} \label{maggiorecoproduct}
\triangle D_{\mu} = D_{\mu}\otimes Z^{-1}+\mathbf{1}\otimes D_{\mu}
 + s D_{\mu} D_{\alpha} \frac{1}{Z^{-1} +1} \otimes D^{\alpha},
\end{equation}
where
\begin{equation}
Z^{-1} = \sqrt{1+sD^2}.
\end{equation}
Relations (\ref{snydercoproduct}) and (\ref{maggiorecoproduct}) appear
 to be equivalent to relations of Ref.\cite{Girelli:2009ii} which give the rules for adding of momenta
and are  obtained by considering a
 momentum addition law on the corresponding momentum space given by a coset.

As indicated earlier, Snyder deformation $(a = 0)$ has a noteworthy property
that, no matter of the realization used, the coproduct for Lorentz
generators is undeformed,
\begin{equation}  
\Delta M_{\mu\nu}=M_{\mu\nu}\otimes1+1\otimes M_{\mu\nu}\,.
\end{equation}
An immediate consequence of this property is that the corresponding
antipodes are also undeformed, 
\begin{equation} \label{antipodes}
S(D_\mu)=-D_\mu\, \qquad S(M_{\mu\nu})=-M_{\mu\nu}\,.
\end{equation}
First one of these relations can alternatively be confirmed by solving
 conditions  (\ref{antipode}) for two special cases of coproducts, 
for the coproduct (\ref{snydercoproduct}) corresponding to realization
 (\ref{snyderrealizationtutu})  or for the coproduct (\ref{maggiorecoproduct})
corresponding to realization (\ref{maggiorerealizationtutu}). In view
 of conditions (\ref{antipode}), both of
 these coproducts
  lead to the same result for antipode and even more, every admissible realization,
for the case of Snyder deformation $(a=0),$ leads to this same result, $S(D_\mu)=-D_\mu.$

\subsection{3. case ($ s=0)$}

 The situation when parameter $\; s \;$ is equal to zero corresponds
 to $\kappa$-deformed space investigated in \cite{KresicJuric:2007nh},\cite{Meljanac:2007xb}.
The  generalized d'Alambertian operator is now given as
\begin{equation}
  \square = \frac{2}{a^2} (1- \sqrt{1- a^2 D^2 }),
\end{equation}
and the general
 form (\ref{3.2}) for the realizations now reduces to
\begin{align}
\x_{\mu} = X_{\mu} \left(-A+\sqrt{1-B}\right)+i(aX)\, D_{\mu},
\end{align}
where $B= a^2 D^2. $
The Lorentz generators can be expressed as
\begin{equation}
M_{\mu\nu}=(\x_{\mu} D_{\nu}-\x_{\nu} D_{\mu}) Z
\end{equation}
and deformed Heisenberg-Weyl algebra (\ref{3.4}) takes on the form
\begin{align}
[D_{\mu},\x_{\nu}] = \eta_{\mu\nu} Z^{-1}+ia_{\mu} D_{\nu}.
\end{align}
In the case of $\kappa$-deformed space, we can also write the
exact result for the coproduct, which in a closed form looks as
\begin{equation}\label{coproductD}
\triangle D_{\mu} = D_{\mu}\otimes Z^{-1}+\mathbf{1}\otimes
D_{\mu}+ia_{\mu} (D_{\alpha} Z)\otimes
D^{\alpha}-\frac{ia_{\mu}}{2} \square\, Z\otimes iaD,
\end{equation}
where the generalized shift operator (\ref{16}) is here specialized to
\begin{align}
Z^{-1}= -iaD+\sqrt{1- a^2 D^2}.
\end{align}
This operator has the following useful properties, with first of them expressing
 the coproduct for the operator $Z,$
\begin{equation}
\triangle Z = Z\otimes Z,
\end{equation}
\begin{align}
\x_{\mu} Z \x_{\nu} = \x_{\nu} Z \x_{\mu}.
\end{align}
The coproducts for Lorentz generators in this case are given in
relation (\ref{coproductmo}) and antipodes for both, translation and
Lorentz generators, can also be expressed in a closed form \cite{Meljanac:2010ps}. 

\subsection{4. case ( Perturbative results up to $a^2$ and $s$)}

In this subsection we shall treat perturbatively the most general case
of deformation $(a,s \neq 0) $ for an arbitrary admissible realization, 
\begin{equation}
  \x_{\mu}=X_{\mu}(-A+f(B))+i(aX) D_{\mu}-(a^2-s) (XD)D_\mu \gamma_2,
\end{equation}
where use will be made of the function
\begin{equation} \label{functiongeneral}
 f(B)= 1 - uB + {\mathcal{O}} (a^3) = 1-u(a^2-s)D^2 + {\mathcal{O}} (a^3),
\end{equation}
expanded in a Taylor series up to second order in deformation parameters
$a$ and $s.$ Here the parameter $u$ plays the role of characterizing
the realization we are working with. Thus, the results for the function $f(B) = \sqrt{1- B}, $
valid up to second order in $a$ and $s,$ will be reproduced for
$u=\frac{1}{2},$ while the results for the function $f(B) = 1, $ valid
within the same order, will be reproduced for $u=0.$
The same procedure which was carried out in section 6 for the function
$f(B) = \sqrt{1- B}, $ results now in the expression
\begin{eqnarray}
P_{\mu}(k,q)  & =  & q_\mu+ k_\mu(1+aq)-a_\mu(kq)-(1-2u)(a^2-s)q_\mu
  (kq)+u(a^2-s)k_\mu q^2 \nonumber \\
  & +  & \frac{1}{2}[k_\mu(ak)-a_\mu k^2](1+aq)+\frac{1}{2}[ a_\mu(ak)-
 k_\mu a^2](kq) \nonumber \\
  & -  & \frac{1}{2}(a^2-s)[(1-4u)k_\mu (kq)+(1-2u)q_\mu k^2]
 +\frac{1}{6}k_\mu[(ak)^2-a^2k^2] \nonumber \\
 & - & \frac{1}{3}(1-3u)(a^2-s)k_\mu k^2 + {\mathcal{O}} (a^3),
\end{eqnarray}
as the solution to Eq.(\ref{4.4}) for the case of function
(\ref{functiongeneral}).
This also gives
\begin{equation} \label{Kmap}
\begin{array}{c}
K_{\mu}=k_\mu[1+\frac{1}{2}(ak)+\frac{1}{6}(ak)^2-\frac{1}{6}a^2k^2-\frac{1}{3}(1-3u)(a^2-s)k^2]-\frac{1}{2}a_\mu
k^2 + {\mathcal{O}} (a^3),
\end{array}
\end{equation}
as an adequate counterpart to the quantity (\ref{4.10}).
The inverse transformation of (\ref{Kmap}) looks as
\begin{eqnarray} \label{knaminusprvu}
K^{-1}_{\mu} (k) &  = & 
k_\mu[1-\frac{1}{2}(ak)+\frac{1}{3}(ak)^2-\frac{1}{12}a^2k^2+\frac{1}{3}(1-3u)(a^2-s)k^2]  \nonumber \\
 & + & \frac{1}{2}a_\mu
k^2-\frac{1}{4}a_\mu(ak)k^2 + {\mathcal{O}} (a^3).
\end{eqnarray}
According to Eq.(\ref{4.12}), these last results immediately yield the relation
\begin{eqnarray} \label{7D}
\D_{\mu}(k,q) &= & q_\mu[1-(1-2u)(a^2-s)(k
q)-\frac{1}{2}(1-2u)(a^2-s)k^2] \nonumber  \\
 &+ & k_\mu[1+(aq)+u(a^2-s)q^2
-\frac{1}{2}a^2(k q)-\frac{1}{2}(1-4u)(a^2-s)(k q)] \nonumber \\
 &+ & a_\mu[(k q)(ak-1)-\frac{1}{2}(aq)k^2] + {\mathcal{O}} (a^3),
\end{eqnarray}
 which in itself comprises a deformed momentum addition rule and a deformed coproduct
for translation generators of Poincar\'{e} algebra. As already pointed
out earlier, due to properties of a pure Snyder deformation $(a=0),$
the coproduct for Lorentz generators in the currently considered case
$(a,s \neq 0, ~~ f(B) = 1 - uB )$ will be the same
as in the corresponding case of a pure $\kappa$-deformation. The same
holds for the antipodes of Lorentz generators.
On the other side, the antipodes for translation generators can be
obtained in a straightforward manner by solving the conditions
(\ref{antipode}), imposed on the function $\D_{\mu}(k,q)$ in Eq.(\ref{7D}).
This gives
\begin{eqnarray}
 S(k_i) & = & -k_i[1+a_0k_0+(a_0k_0)^2-\frac{1}{2}a_0^2k^2] + {\mathcal{O}} (a^3),  \nonumber \\
  S(k_0) & = & -k_0(1-a_0^2 \sum_{i=1}^{n-1} k_i^2)-a_0
  \sum_{i=1}^{n-1} k_i^2 + {\mathcal{O}} (a^3),
\end{eqnarray}
with the property
\begin{equation}
 {(S(k))}^2 = - {(S(k_0))}^2  + \sum_{i=1}^{n-1}{(S(k_i))}^2  = k^2,
\end{equation}
which is, of course, valid within the second order in $a$ and $s$.


We briefly show how the results obtained  can be used
to construct a field theory for free, as well as for interacting field
theory. As for toy model, we consider a scalar field theory with mass and cubic
interaction terms.

Having in mind Eqs.(\ref{4.1}) and (\ref{4.2}), we have the following associations,
\begin{equation} \label{dodatak1}
 e^{iK^{-1}(k) \x} \triangleright 1 = ~ e^{ikX}
\end{equation}
and
\begin{equation} \label{dodatak2}
 e^{iK^{-1}(k)\x} \triangleright e^{iqX} ~ = ~
 e^{iP_{\mu}(K^{-1}(k),q)X^{\mu}} ~ = ~ e^{i {{\mathcal{D}}_{\mu} (k,q)}X^{\mu}},
\end{equation}
as well as
\begin{equation} \label{dodatak3}
 D_{\mu} e^{iK^{-1}(k) \x} \triangleright 1 ~ = ~ D_{\mu} e^{ikX} ~ = ~
  ik_{\mu} e^{ikX}.
\end{equation}
From the 
definition of the star product (\ref{starproductdefinition}), we can
write
\begin{eqnarray} \label{dodatak4}
\hat{\phi}(\x)\hat{\phi}(\x) \triangleright
1  & = &  \phi (X) \star \phi(X),  \\
\hat{\phi}(\x) \hat{\phi}(\x) \hat{\phi}(\x) \triangleright 1
  & = &  \phi (X) \star (\phi(X) \star \phi(X)).
\end{eqnarray}
We also assume that functions in NC coordinates and functions in
commutative coordinates have the following
Fourier transforms, respectively,
\begin{eqnarray} \label{dodatak5}
\hat{\phi}(\x)  & = &  \int \frac{d^n k}{{(2\pi)}^n} \hat{\phi} (k) e^{iK^{-1}(k) \x}, \\
\phi (X)  & = &  \int \frac{d^n k}{{(2\pi)}^n} \hat{\phi} (k) e^{ik X},
\end{eqnarray}
so that they can mutually be related through the association $\hat{\phi}(\x) \triangleright 1 = \phi (X). $
 That  this is really the case, can
easily be inferred from Eq.(\ref{dodatak1}).

The action for interacting massive scalar field on spacetime (\ref{2.1})
can then be obtained by the projection on the unit element as follows,

\begin{align} \label{dodatak7}
 S[\phi] & = \int d^n X \left( \frac{1}{2} ( D_{\mu}\hat{\phi} D^{\mu}\hat{\phi} +
 m^2 {\hat{\phi}}^2)  + \frac{\xi}{3!} {\hat{\phi}}^3\right)
 \triangleright 1  \nonumber \\
 & = \frac{1}{2}\int d^n X ~ (D_{\mu}\phi) \star (D^{\mu}\phi) +
 \frac{m^2}{2}\int d^n X ~ \phi \star \phi 
    + \frac{\xi}{3!} \int d^n X ~ \phi \star (\phi \star \phi). 
\end{align}
By making use of Fourier transforms and associations given in
Eqs.(\ref{dodatak2}) and (\ref{dodatak3}), various terms appearing in (\ref{dodatak7}) are calculated as follows
\begin{eqnarray} \label{dodatak6}
D_{\mu}\hat{\phi}(\x) D^{\mu}\hat{\phi}(\x) \triangleright
1  & = &  D_{\mu}\hat{\phi}(\x) \triangleright \left(
  D^{\mu}\hat{\phi}(\x) \triangleright 1 \right) = (D_{\mu} \phi) \star
(D^{\mu} \phi)  \nonumber \\
 & = & \int \frac{d^n k}{{(2\pi)}^n} \int \frac{d^n q}{{(2\pi)}^n} \hat{\phi}(k) \hat{\phi}(q) 
 \left( q^2  - q {\mathcal{D}} \right)  e^{i {{\mathcal{D}} (k,q)}X},
\end{eqnarray}
\begin{eqnarray} \label{dodatak6b}
{\hat{\phi}(\x)}^2  \triangleright
1  & = &  \int \frac{d^n k}{{(2\pi)}^n} \int \frac{d^n q}{{(2\pi)}^n} \hat{\phi}(k) \hat{\phi}(q) 
      e^{iK^{-1}(k)\x} \triangleright \left(  e^{iK^{-1}(q)\x} \triangleright 1 \right) \nonumber \\
& = & \int \frac{d^n k}{{(2\pi)}^n} \int \frac{d^n q}{{(2\pi)}^n} \hat{\phi}(k) \hat{\phi}(q) 
      e^{i {\mathcal{D}}_{\mu} (k,q) X^{\mu}},
\end{eqnarray}
\begin{eqnarray} \label{dodatak6a}
{\hat{\phi}(\x)}^3  \triangleright
1  & = &  \int \frac{d^n p}{{(2\pi)}^n} \int \frac{d^n k}{{(2\pi)}^n} \int \frac{d^n q}{{(2\pi)}^n} \hat{\phi}(p) \hat{\phi}(k) \hat{\phi}(q) 
     e^{iK^{-1}(p)\x}  \triangleright \left( e^{iK^{-1}(k)\x} \triangleright
    \left(  e^{iK^{-1}(q)\x} \triangleright 1 \right) \right)
  \nonumber \\
 & = &  \int \frac{d^n p}{{(2\pi)}^n} \int \frac{d^n k}{{(2\pi)}^n} \int \frac{d^n q}{{(2\pi)}^n} \hat{\phi}(p) \hat{\phi}(k) \hat{\phi}(q) 
     e^{iK^{-1}(p)\x}  \triangleright e^{iP_{\mu}(K^{-1}(k),q)X^{\mu}}
     \nonumber \\
 & = &  \int \frac{d^n p}{{(2\pi)}^n} \int \frac{d^n k}{{(2\pi)}^n} \int \frac{d^n q}{{(2\pi)}^n} \hat{\phi}(p) \hat{\phi}(k) \hat{\phi}(q) 
      e^{i P_{\mu}\left(K^{-1}(p), P(K^{-1}(k),q)\right)X^{\mu}}
      \nonumber \\
& = &  \int \frac{d^n p}{{(2\pi)}^n} \int \frac{d^n k}{{(2\pi)}^n} \int \frac{d^n q}{{(2\pi)}^n} \hat{\phi}(p) \hat{\phi}(k) \hat{\phi}(q) 
      e^{i {\mathcal{D}}_{\mu} \left( p, {\mathcal{D}}(k,q) \right)X^{\mu}},
\end{eqnarray}
where $K^{-1}(k), {\mathcal{D}} (k,q)$ are given in
 Eqs.(\ref{knaminusprvu}) and (\ref{7D}), respectively.
 Note the appearance of nested terms like ${\mathcal{D}}_{\mu} \left(
 p, {\mathcal{D}}(k,q) \right).$ These can be calculated by applying
 Eq.(\ref{7D}) repeatedly. As pointed out before, relations (\ref{knaminusprvu}) and (\ref{7D})
 represent the most general case of deformation ($a,s \neq 0$ and
 parameter $u$ is arbitrary) and
 they are obtained perturbatively up to second order in deformation parameters.
    In obtaining Eqs.(\ref{dodatak6}),(\ref{dodatak6b}) and (\ref{dodatak6a}), we have
 used relation (\ref{4.12}) together with
 relations (\ref{dodatak1}),(\ref{dodatak2}) and (\ref{dodatak3}).



The action (\ref{dodatak7}) can be expanded by making use of
Eq.(\ref{7D}) and the above results.
For transparency, we look at somewhat simpler case when $a=0$. Then
the expansion up to linear order in parameter $s$
leads to a standard action with the additional
correction terms,
\begin{align} 
  S[\phi] & = \int d^n X ~ {\mathcal{L}}(\phi, D_{\mu}\phi, D_{\mu} D_{\nu} \phi) \nonumber \\
 & = \frac{1}{2}\int d^n X ~ (D_{\mu}\phi) (D^{\mu}\phi) +
 \frac{m^2}{2}\int d^n X ~ \phi^{2} + \frac{\xi}{3!} \int d^n X ~ \phi^{3}  \nonumber \\
 & + \frac{s}{4} \int d^n X ~
 (X^{\mu} D_{\mu} D_{\lambda} \phi) D^{\lambda} D^{2} \phi
  + \frac{s}{4} \int d^n X ~ (X^{\mu} D_{\mu} D_{\nu} D_{\lambda}
 \phi) D^{\nu} D^{\lambda} \phi \nonumber \\
 & + s\frac{m^2}{4} \int d^n X ~ (X^{\mu} D_{\mu} \phi) D^{2} \phi 
 + s\frac{m^2}{4} \int d^n X ~ (X^{\mu} D_{\mu}D_{\nu} \phi) D^{\nu} \phi \nonumber \\
 & + s\frac{\xi}{4} \int d^n X ~ \phi (X^{\mu} D_{\mu} \phi) D^{2} \phi 
 + s\frac{\xi}{4} \int d^n X ~ \phi (X^{\mu} D_{\mu}D_{\nu} \phi)
 D^{\nu} \phi \nonumber \\
& + s\frac{\xi}{3!} \int d^n X ~  (X^{\mu} D_{\mu} \phi) (D_{\nu} \phi)
 D^{\nu} \phi  + {\mathcal{O}}(s^2).
\end{align}


\section{Conclusion}

   In summary, the focus of our analysis was directed toward
   $\kappa$-Snyder deformation of Minkowski spacetime. This
   deformation is of a Lie algebra type and it includes features of both, a
   pure $\kappa$-deformation and of pure Snyder deformation, at the
   same time. It in fact interpolates between the two in a smooth way,
   broadening a possible range of deformations, thus making features
   resulting from that extension more likely to  correspond to and to fit within
   the scope of what is really
   happening at the Planck scale level. The analysis is further made
   of the impact that these deformations have on the Hopf algebraic structure
   of the symmetry algebra underlying Minkowski space.
  Particularly, the nature of the relationship between
   algebraic structures, the original, undeformed one and the one
   resulting from a deformation of an underlying spacetime geometry is considered.
  Although these algebraic structures are not isomorphic to each
   other, we have however shown that this situation can be overcome by introducing a notion of module
   for the corresponding universal enveloping algebras.
  A special thing about introducing a module is that when we take a
   unit element in the module and project the enveloping algebras 
    of the deformed and undeformed Heisenberg algebras
   to the unit element in the module, respectively, we arrive at the conclusion that
   the resulting structures are isomorphic
   to each other. This path appears to be the right way in which the
   isomorphism can again be established, even in the most general case
   of deformation, when both deformation parameters are different from
   zero $(a,s \neq 0) $. We have further investigated the way in which 
   a  construction of tensors and
    invariants, in terms of NC coordinates, should be modified in
   order for it to be
   compatible with Lorentz symmetry and to avoid all inconsistencies
   that could possibly arise on the way. 
Deformations that have been studied are further found to completely fit within the 
framework of a quantum Hopf algebra. They are characterized by
 the common feature that the algebraic sector
     of the Hopf algebra, which
     is described by the Poincar\'{e} algebra, is  undeformed, while, on the
     other side, the corresponding coalgebraic sector is
     affected by deformations.
   Deformation of the coalgebra manifests in a form of   
     having the modified  coproducts for Poincar\'{e} generators, which in turn
     tell us to which extent
   the corresponding Leibniz rules are deformed in comparison to standard Leibniz rules. Since the
   coproduct is related to a
 star  product, we were also able to
    write down
    how star product looks like for NC spaces characterized by the
    general class of deformations of type (\ref{2.1}).
     We have also found many different  classes of realizations of NC space (\ref{2.1})
   and specialized obtained results to some specific cases of
    particular interest, including the perturbative analysis of the most general
    case of realization, valid up to second order in deformation parameters.

 The point that might be important to emphasize is that    
 the realizations that we have been working with in this paper are not hermitian
    ones, but they could be made hermitian. This procedure of hermitization has led in
    the case of pure $\kappa$-space to a
  very important result satisfied by the star product corresponding to
    hermitian realization, namely, that under the integration sign,
    star product can be replaced by the ordinary multiplication
    operation \cite{Meljanac:2010ps}. Consequences that
    hermitization process has for the trace and cyclic properties of an
    invariant integral have also been discussed for $\kappa$-space \cite{Meljanac:2010ps}.
   We would expect a similar kind of results emerging in $\kappa$-Snyder
    space as well,  if the similar
    process of hermitization was carried out there. These matters and
 particularly the issue of the trace property of an invariant integral
 defined on noncommutative spacetime are highly relevant
 in building field theories and  gauge theories on these noncommutative manifolds. 
It hence remains challenging to address the problem of invariant
 integration on $\kappa$-Snyder deformed space and 
  to construct a scalar
 field theory, following the work that was previously done
 in the context of $\kappa$-deformation \cite{Daszkiewicz:2007az},\cite{Freidel:2006gc},\cite{Meljanac:2010ps},\cite{Meljanac:2007xb},
as well as in the context of Snyder deformation \cite{Girelli:2009ii},\cite{Girelli:2010wi},\cite{Battisti:2010sr}.

   There is also a vide range of physicaly appealing questions which 
      could be expected to originate from the modified geometry
      at the Planck scale, which reveals itself through a
      noncommutativity of spacetime coordinates. Some of these questions are
      related to investigations of the effects that noncommutativity
      has on dispersion relations  \cite{Magueijo:2001cr},\cite{AmelinoCamelia:2009pg}, black hole horizons \cite{Kim:2007nx},
     and  Casimir energy \cite{Kim:2007mb}, the issues that have already
      been analysed in the context of
       $\kappa$-type noncommutativity.
  However, since in our approach Poincar\'{e} algebra is undeformed,
    dispersion relation will also be undeformed, being in line with
    the standard dispersion relation, $P^2 + m^2 = 0.$
  In case we treated derivative as not being vector-like, we would get
      modified dispersion relations.
  In order to formulate field theory for the case of $\kappa$-Snyder deformation
 and to investigate the impacts that deformation has on  particle
      statistics, as well as on certain important physical
      properties such as Lorentz and  CPT invariance,
 it is necessary to find
 Drinfeld twist \cite{Borowiec:2004xj},\cite{borowiec},\cite{Bu:2006dm},\cite{Govindarajan:2008qa},
   twisted flip operator \cite{Govindarajan:2008qa},\cite{Young:2007ag},\cite{gov}
 and $ R$-matrix \cite{Young:2008zm},\cite{gov}, that are relevant on spacetime
    with  $\kappa$-Snyder deformation. The issue of proper
      construction of
differential forms \cite{Meljanac:2008pn},\cite{Kim:2008mp} would also
      be of significant importance. 
   The most of these issues will be addressed in  the forthcoming papers, particularly the
   issues related to field theory
   for scalar fields 
 and its twisted statistics properties, as a
   natural continuation of our investigations put forward in previous
   papers \cite{Govindarajan:2008qa},\cite{gov}. At the Planck scale it is also not clear
      at all if the Lorentz invariance  has
      remained intact, or is it violeted. From the theoretical point
      of view, the problem of Lorentz invariance, in conditions
      regulated by  $\kappa$-Snyder deformation, can also be  accessed 
    by assuming that the deformation parameter $a,$ instead of being fixed, transforms as a 
      $n$-vector under Lorentz algebra, which is a kind of approach already
      considered in \cite{Meljanac:2010ps}.

{\it{Acknowledgment}}.
   We are grateful to Marco Valerio Battisti for critical reading of
   the manuscript and for useful comments.
    We also thank Domagoj Kovacevic for valuable comments.
     This work was supported by the Ministry of Science and Technology
    of the Republic of Croatia under contract No. 098-0000000-2865.


\end{document}